\documentclass[manuscript]{acmart}

\usepackage{color}
\usepackage{listings}

\AtBeginDocument{%
  \providecommand\BibTeX{{%
    \normalfont B\kern-0.5em{\scshape i\kern-0.25em b}\kern-0.8em\TeX}}}

\copyrightyear{2021} 
\acmYear{2021} 
\setcopyright{acmlicensed}\acmConference[ARES 2021]{The 16th International Conference on Availability, Reliability and Security}{August 17--20, 2021}{Vienna, Austria}
\acmBooktitle{The 16th International Conference on Availability, Reliability and Security (ARES 2021), August 17--20, 2021, Vienna, Austria}
\acmPrice{15.00}
\acmDOI{10.1145/3465481.3470079}
\acmISBN{978-1-4503-9051-4/21/08}

\lstdefinestyle{xmlStyle}{
  belowcaptionskip=1\baselineskip,
  breaklines=true,
  frame=single,
  rulecolor=\color{black}, 
  xleftmargin=0.1cm,   
  xrightmargin=0.1cm,  
  captionpos=b, 
  language=C,
  showstringspaces=false,
  basicstyle=\footnotesize\ttfamily,
  keywordstyle=\bfseries\color{green!40!black},
  commentstyle=\itshape\color{purple!40!black},
  stringstyle=\color{blue},
  morekeywords={beginning,property,embedded_functions,event,boolean_expression,if_satisfied,update,drop,uint64_t},
}

\begin{document}

\title{5Greplay: a 5G Network Traffic Fuzzer - Application to Attack Injection}



\author{Zujany \textsc{Salazar}, Huu Nghia \textsc{Nguyen}, Wissam \textsc{Mallouli},
    Ana R. \textsc{Cavalli}, Edgardo \textsc{Montes de Oca}}

\authornotemark[1]
\email{firstname.lastname@montimage.com}
\affiliation{%
  \institution{Montimage}
  \streetaddress{37 rue Bobillot}
  \city{Paris}
  \country{France}
  \postcode{75013}
}


\renewcommand{\shortauthors}{Salazar et al.}

\begin{abstract}
The fifth generation of mobile broadband is more than just an evolution to provide more mobile bandwidth, massive machine-type communications, and ultra-reliable and low-latency communications. It relies on a complex, dynamic and heterogeneous environment that implies addressing numerous testing and security challenges. In this paper we present \href{http://5greplay.org}{5Greplay}, an open-source 5G network traffic fuzzer that enables the evaluation of 5G components by replaying and modifying 5G network traffic by creating and injecting network scenarios into a target that can be a 5G core service (e.g., AMF, SMF) or a RAN network (e.g., gNodeB). The tool provides the ability to alter network packets online or offline in both control and data planes in a very flexible manner. The experimental evaluation conducted against open-source based 5G platforms, showed that the target services accept traffic being altered by the tool, and that it can reach up to 9.56 Gbps using only 1 processor core to replay 5G traffic.
\end{abstract}


\maketitle

\section{Introduction}

The evolution of 5G mobile networks towards a service-based architecture (SBA) comes with the emergence of numerous new testing challenges and objectives. First, 5G deployment introduces a brand-new set of technologies, such as, the network function softwarization enabled by the software defined networking (SDN) and network functions virtualization (NFV), Mobile edge computing (MEC), and Network Slicing (NS). These require to be tested from a functional point of view; but, also from a non-functional point of view in order to determine the sanity of the system based on indicators such as data throughput performance, latency, scalability, robustness, etc. Finally, 5G SBA introduces new cybersecurity threats, with some of the previously adopted security and privacy mechanisms becoming ineffective or not applicable in 5G due to the changes in the architecture and the advent of new services  \cite{ahmad}. This requires the creation of new sets of security test cases and tools specifically targeting 5G security concerns.

The 3GPP standardization organism has proposed a wide set of test cases to check functional and non-functional requirements in 5G network products. For example, in the TS 33.117 catalogue\cite{ts33117} of general 5G security assurance requirements, they propose a group of test cases to verify if network products providing externally reachable services are robust against unexpected inputs. The target of these tests are the 5G protocol stacks (e.g., Diameter stack). Other specifications, such as the TS 33.512\cite{ts33512}, concerning the security assurance of the Access and Mobility management Function (AMF), provides test cases to verify that this 5G component is properly protected against specific vulnerabilities. \par

Regarding security testing, 5G issues have been the subject of numerous studies. Standardization organisms list collections of threats and vulnerabilities \cite{etsi_nfv}, also investigated by  academia \cite{basin2018,ahmad}, and Industrial researchers \cite{ericsson_2018}\cite{positivetech_2021}. Several sources have studied the problematic of replay attacks in  5G networks, that could expose the system to Man-in-the-Middle (MiTM) or Denial-of-Service (DoS) attacks. Technology reports identify the possibility of malicious actors performing a DoS or MiTM attack by replaying Packet Forwarding Control Protocol (PFCP) messages that manage GPRS Tunnelling Protocol (GTP) tunnels. The ENISA organisation\cite{enisa_2021} alerts that an AMF can be vulnerable to replay attacks of Non-Access Stratum (NAS) signaling messages between the UE and AMF on the N1 interface. Moreover, ENISA adds that the security of a network slice could be compromised if an attacker spoofs a genuine network manager by obtaining access to an insecure network management interface, or just by replaying or modifying a valid message.\par

In order to overcome these issues, numerous research has been done in the matter of  detection and prediction of 5G cyber-attacks. Several Intrusion Detection Systems (IDSs) have been proposed \cite{ monge2019,moudoud2021}. Nevertheless, to the best of our knowledge, there is a lack of open-source solutions that enable to manually create or edit existing 5G network protocol packets and injecting them in a network, allowing to easily test the proposed detection schemes. \par

Therefore, besides the significant amount of functional and non-functional proposed test cases, as well as the previously mentioned collections of 5G network threats and vulnerabilities and intrusion detection approaches, the testing of 5G network components and IDSs remains a challenge. This is in part due to the lack of publicly available labeled data sets containing realistic user behavior and up-to-date attack scenarios \cite{RING2019156}. 
 
Open-source tools, such a Tcpreplay, aim solving this issue by enabling to replay malicious traffic patterns on IDSs. More recent versions of the tool include the capability to replay traffic on web servers. Within the Tcpreplay suite, Tcpwrite allows editing, creating and replaying PCAP files in a network. However, it targets modifying IP, TCP and UDP attributes and fields. Other packet manipulation solutions such as Scapy are also not 5G-oriented. 

Here, we propose 5Greplay, an open-source solution to perform fuzz testing of 5G networks. 5Greplay aims to facilitate the testing process of 5G virtual network functions and IDSs by allowing to forward network packets from one network interface card (NIC) to another with or without modifications. The tool supports the implementation of test cases by letting the users create specific scenarios using PCAP files that contain conventional 5G network traffic and execute them on a target network.

The preliminary experimentation shows that the traffic generated by the tool is correct with respect to the protocols' standardization and is accepted by open-source 5G frameworks, such as, open5GS, free5GC. The tool gives users the ability to alter 5G packets in a very flexible way to perform  tests and attack scenarios. The experimental results  also show the scalability of the tool to replay very high-bandwidth traffic.

The rest of the paper is structured as follows. Section~\ref{sec:architecutre} presents the tool's architecture and functionalities. It also presents possible use cases for performing 5G functional or non-functional tests.  Section 3 presents technical details about the tool's implementation and the background information about the SDK library and rule engine used by the tool. Section~\ref{sec:experimentation} presents three experimental scenarios used to evaluate the tool's efficiency. And finally, the paper ends in Section~\ref{sec:conclusion} with the conclusion and possible future work.

\section{5Greplay General Overview}
\label{sec:architecutre}

\begin{figure}[ht]
    \centering
    \includegraphics[width=10cm]{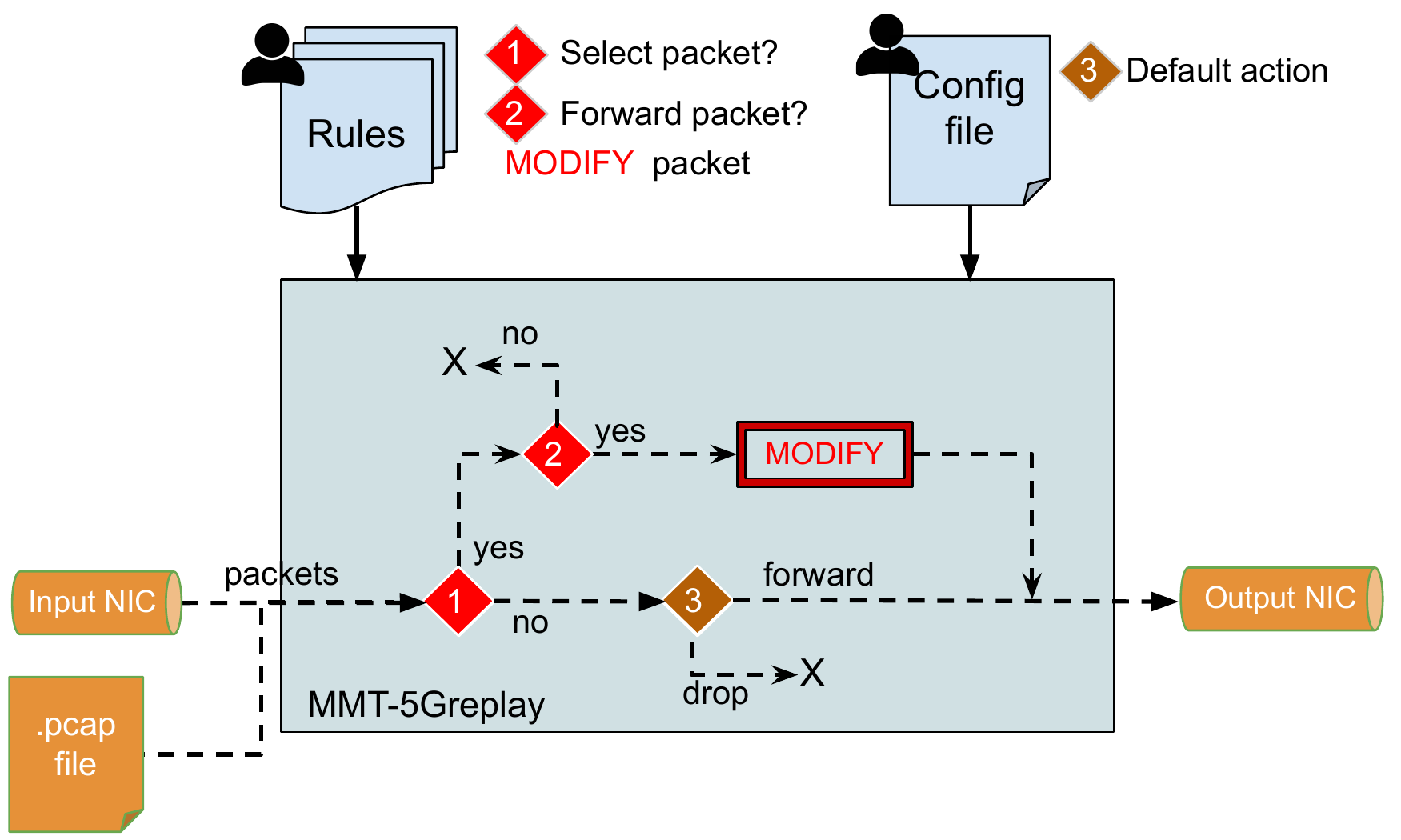}
    \caption{5Greplay main process. Incoming network packets are filtered according to predefined rules (see Section \ref{rules}) that determine which packets will be modified, forwarded, or dropped before being sent to the output NIC}
    \label{fig:arch}
\end{figure}

\href{http://5greplay.org}{5Greplay} is an open-source solution entirely developed by the authors that allows forwarding network packets from one network interface card to another with or without modification. It can be considered as a one-way bridge between the input NIC and the output one. It can also take as input pre-captured packets that are saved in a PCAP-format file.\par
Its behavior is controlled by user defined rules and completed by a configuration file. The user defined rules allow explicitly indicating which packets can be passed through the bridge and how a packet is to be modified in the bridge. The configuration file allows specifying the default actions to be applied on the packets that are not managed by the rules, i.e., if they should be forwarded or not. 5Greplay's global architecture is depicted in Figure~\ref{fig:arch}. \par

For each arriving packet, 5Greplay classifies the packet to identify its protocol, extracts the protocol attributes that are used in the rules, and checks whether each rule in the set of rules is satisfied or not. If no rules are satisfied, 5Greplay applies the default action that is defined in the configuration file. The default action is either: \texttt{FORWARD} to forward the packet to the output NIC without any modification, or \texttt{DROP} to drop the packet. Otherwise, if the packet satisfies a rule, 5Greplay will apply the action defined by the rule (see Section \ref{rules}). If the action is \texttt{FORWARD}, then 5Greplay modifies the packet as defined in the rule before sending it to the output NIC. \par

In the following sections, we give technical details of 5G architecture. Section \ref{rules} describes the syntax of the rules that determine the behaviour of the tool, and Section \ref{usecases} shows different configurations in which the tool can be use. System requirements for running 5Greplay, usage instructions, and more detailed technical information, as well as the tool itself, can be found in the following Montimage dedicated url: \url{http://5greplay.org}.

\subsection{Rules} \label{rules}

5Greplay rules allow users to select specific packets in an incoming stream by using Deep Packet Inspection (DPI) techniques to classify and extract protocol attributes. When defining a replay rule, users must indicate the following three elements in the rule: 1) which packet will be processed, 2) which action will be applied, and 3) how to modify the~packet. \par

The rules are specified in XML format, Listing~\ref{lst:rule1} presents the rule used in the scenario 2 (see Section~\ref{sec:scenario-smc-replay-attack}). The objective of this rule is to created malformed Next Generation Application Protocol (NGAP) packets and replay them on the targeted 5G core network. First, the user defines the desired actions to take for the filtered packets in the \texttt{if\_satisfied} attribute. In this example, the action is to change the Stream Control Transmission Protocol (SCTP) protocol identifier from 60, that corresponds to the NGAP protocol, to 0. Then, the user defines the conditions to filter the packets on which the actions will be performed. These conditions are listed as events that happen one after the other following the time limits specified by \texttt{delay\_units}, \texttt{delay\_min}, \texttt{delay\_max} fields. In the example, the time between events must be more than 0ms and less than 1ms. Finally, the events are defined. Each event has an identifier \texttt{event\_id}, a description, and a condition expressed in the \texttt{ boolean\_expression} field. In the example, there are two events, in other words, two filtering conditions. First, there must be a packet with a NAS message type equal to 93, corresponding to a NAS Security mode Command message. Then, if less than 1ms later there is a packet with NAS Security type equal to 4, corresponding to a NAS Security mode Complete message, this packet will be modified as specified by the \texttt{if\_satisfied} attribute and forwarded to the address indicated in the 5Greplay configuration file.\par

5Greplay parses most Internet protocols and applications. Regarding 5G dedicated protocols, the tool currently enables filtering 5G packets of Non-Access-Stratum protocol for 5G System (NAS-5G), NG Application Protocol (NGAP), GPRS Tunnelling Protocol version 2 (GTPv2), Stream Control Transmission Protocol (SCTP) and DIAMETER protocols. To perform the fuzzing operators defined in Table \ref{tab:operators}, 5Greplay can operate over SCTP, NAS-5G and NGAP protocols. The tool incorporates a plugin architecture for the addition of new protocols and data structures; furthermore, if a specific test scenario requires parsing a protocol that is not currently supported, the addition of it is possible in a simple manner. 

\begin{lstlisting}[caption={Forwarding a NAS security mode Complete message that answers to the NAS security mode Command message}, style=xmlStyle, label=lst:rule1]
<property value="THEN"  delay_units="ms" delay_min="0" delay_max="1" property_id="100" type_property="FORWARD" description="Forwarding NAS security mode COMPLETE that answers to NAS security mode COMMAND" if_satisfied="#update(sctp_data.data_ppid, 0)">
    <event event_id="1" description="NAS Security mode COMMAND"
           boolean_expression="(nas_5g.message_type == 93)"/>
    <event event_id="2" description="NAS Security mode COMPLETE"
           boolean_expression="(nas_5g.security_type == 4)"/>
</property>
\end{lstlisting}

\subsection{Use cases} \label{usecases}

\subsubsection{Classical Traffic Replay}
5Greplay is intended for providing a complementary solution to existing tools that support 5G network traffic.
The use cases of 5Greplay can be classified into two categories, offline and online, depending if the input packets are coming from a pre-captured PCAP file or live streaming.
When inputting a PCAP file, it can be considered as a classical traffic replay tool (e.g., Tcpreplay and Tomahawk). However, it differs from the existing tools, which usually replay pre-captured packets without altering them or only allow modifying them at the IP layer in a predefined manner, such as, Tcpreplay that allows increasing the IP source and destination fields after each iteration. On the other hand, 5Greplay focuses on the modification of 5G protocols, such as, NAS-5G, NGAP. Moreover, the tool empowers the users to be able to alter arbitrary attributes in an arbitrary way.

\subsubsection{Service Chains}
\begin{figure}
    \centering
    \includegraphics[width=8cm]{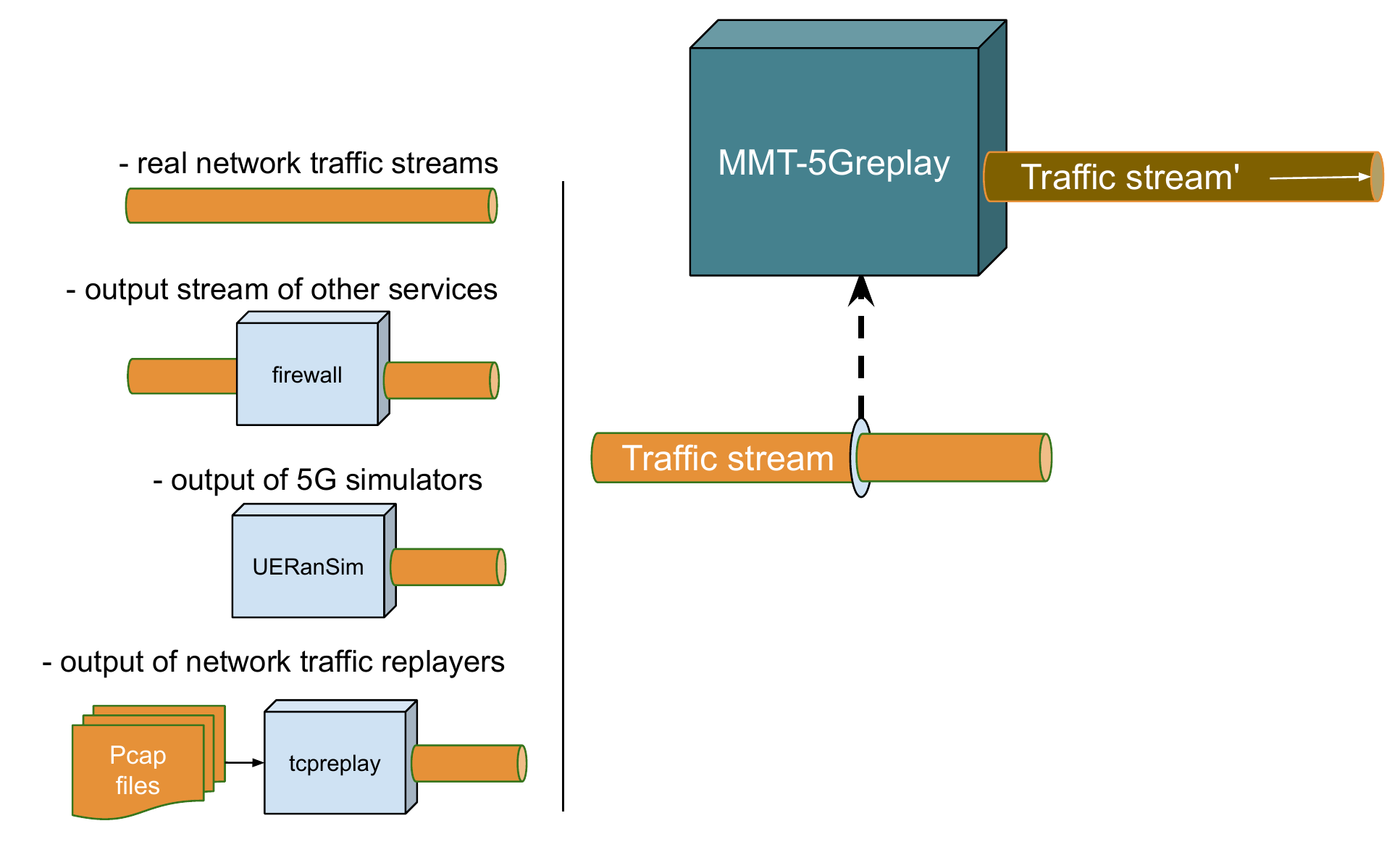}
    \caption{5Greplay in Service chaining}
    \label{fig:usecase}
\end{figure}

5Greplay is able to alter 5G packets in real-time and can be easily combined with other tools. Its input is a network traffic stream that may be the output of other tools or services. We list some possible combinations of 5Greplay with existing tools in Figure \ref{fig:usecase}. As shown, 5Greplay can be placed after another network replay tool, such as, Tcpreplay to be able to provide more capabilities for altering packets. For example, the TCP/IP layer can be managed by Tcpreplay, while our tool can take care of managing the 5G protocol layers.

The input of 5Greplay can also be a real network traffic stream, for instance in the Radio access network (RAN) or Core networks. The tool will select some interesting packets to be altered, then output them to the output stream \texttt{traffic\_stream'} that is then merged into the original \texttt{traffic\_stream} to allow injecting the new altered packets into the legitimate network traffic. The same scenario can be applied on an experimental environment in which the real 5G traffic stream is generated by a simulator. In  our evaluation experiments, we use UERANSIM to simulate the 5G-SA (5G Stand Alone) RAN network since there is yet no open-source solution available for the 5G-SA RAN network.


\section{Background}  \label{background}




\subsection{Fuzz testing and mutant operators}

Mutation testing is a software testing method that aims detecting faults by generating numerous similar test cases. It does this by changing a program \textit{P} under test according to specific rules \textit{R} to obtain a new syntactically correct program~\textit{M}. \textit{M} is called a \textit{mutant} of \textit{P}, and \textit{R} is referred to as the \textit{mutant operator} \cite{WONG1995185}. \par

5Greplay aims performing fuzz testing, a type of mutation testing that injects invalid, unexpected, or random inputs to evaluate the response of a test target, in this case 5G virtual network functions, IDSs, 5G applications, etc. It generates \textit{mutants} of the network traffic by using \textit{mutant operators}, in order to perform specified security and functional tests on a system, as well as generating many new test variants. \par

Let \textit{P} denote a 5G network packet (e.g., the security mode command message that the AMF sends to the UE to initiate the SMC procedure) in a PCAP file or a specific real-time flow of network packets, Table \ref{tab:operators} depicts the simplest operators that we consider necessary to mutate the network traffic, and that can be combined in order to generate complex mutants of the original traffic. 

\begin{table}[!ht]
    \centering
    \begin{tabular}{|l|l|}
    \hline
        Atomic operator &  Description \\ \hline  \hline
        DEL\_PKT(\textit{P}) &  Delete a packet  \\ \hline
        CH\_ATTR(\textit{P}) &  Change a specific attribute on the header of a network protocol message    \\ \hline
        ORD(\textit{P1, P2})* & Exchange the order of two consecutive packets \\ \hline
        DUP\_PKT(\textit{P}) &  Duplicate packet \\ \hline
    \end{tabular}
    \caption{5Greplay atomic operators. *Currently not implemented}
    \label{tab:operators}
\end{table}

\subsection{DPI - Deep Packet Inspection}
DPI is an advanced technique used for classifying network traffic~\cite{Valenti2013}. It aims at replacing the first generation of port-based classification methods that traditionally examine network packets by searching information in the packets' headers. The transport protocol and application ports are usually sufficient to identify the application protocol. With the emergence of port-independent, peer-to-peer and encrypted protocols, the task of identifying application protocols has become increasingly challenging. DPI examines the contents of packets passing through a given checkpoint and a set of signatures not only in the headers but also in the payloads in order to classify the packets more accurately. This is possible only if the payload is not encrypted, can be decrypted, or certain data can be extracted, such as statistical data, and analysed, for example using machine learning techniques, that can be used to categorize the packets or flows. 

5Greplay relies on MMT-DPI, a C library that implements DPI techniques to analyse network traffic in order to classify the network protocol or application a packet belongs to, and to extract network and application based events, such as, protocols field values, network and application QoS parameters, and KPIs. MMT-DPI implements a plugin architecture to allow easily adding new protocols and analysis techniques. It provides a set of public APIs for integration with third party probes. The library has been developed by the authors and it is Montimage property, but the necessary part of it will be part of the 5Greplay open-source project.

\section{Experimental Evaluation}
\label{sec:experimentation}

The main objectives of the preliminary experimentation are: i) to determine whether the altered packets generated by the tool are accepted by 5G services/components, ii) to verify that the traffic injection works correctly, and iii) to test the scalability of the tool. The following section will introduce four of the selected testing scenarios to evaluate these objectives. The scenarios are focused on altering 5G traffic in the N1 and N2 interfaces between the RAN and Core~networks. \par

The fist scenario is a simple offline modification of a PCAP file. The second scenario aims modifying online packets. The third scenario intends creating a replay attack against a simulated AMF. These scenarios test the traffic injection capability of 5Greplay, as well as check if the altered packets are accepted by an Intrusion Detection System, and the 5G core simulators. The last scenario is dedicated to evaluating the scalability of the tool by generating high-bandwidth traffic and, thus, to be able to perform a DoS attack (or stress testing) against a target component/service.

\subsection{Scenario 1: Modifying and injecting network traffic offline}
\begin{figure}
    \centering
    \includegraphics[width=6cm]{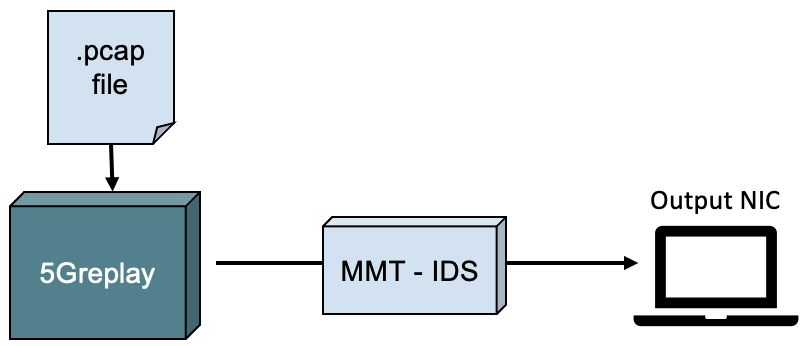}
    \caption{Offline Example -- Sending modified network traffic between two virtual machines}
    \label{fig:scenario1}
\end{figure}

\subsubsection{Description}
This use case scenario evaluates the capability of 5Greplay to modify packets from a pre-existing PCAP file and replaying it on a specific NIC. It tests if the altered packets can be processed by an Intrusion Detection System.

\subsubsection{Testbed Setup}

To perform this scenario, we designed the 5Greplay rule depicted in Listing~\ref{lst:rule-modif-packet}, to be applied on a PCAP file containing 5G-SA traffic. Within an UE authentication message exchange, the rule modifies the UE RAN identifier in any NGAP authentication response, i.e., \texttt{ngap.procedure\_code == 46}, in order to make it different from the one in the NGAP authentication request. The rule increases the UE identifier by 100. Then, we replayed this packet together with the rest of the traffic between two virtual machines. We monitored this activity with the Montimage Monitoring Tool (MMT), an IDS that triggers an alert if in the same SCTP flow, an Authentication Request message and an Authentication Response, contains the same UE identifier for the AMF, but different UE identifier for the RAN, indicating a suspicious activity. We depict this scenario in the Figure~\ref{fig:scenario1}.  \par

\begin{lstlisting}[caption={5G packets alternating using 5Greplay}, style=xmlStyle, label=lst:rule-modif-packet]
<property property_id="101" type_property="FORWARD" description="Increase UE RAN Id in authentication response NGAP messages" if_satisfied="#update(ngap.ran_ue_id, (ngap.ran_ue_id.1 + 100)">
    <event event_id="1" description="Authentication response, Uplink NAS Transport"
        boolean_expression="((ngap.procedure_code == 46) &amp;&amp;  (nas_5g.message_type == 87))"/>
</property>
\end{lstlisting}

\subsubsection{Result analysis}
\begin{figure}
    \centering
    \includegraphics[width=10cm]{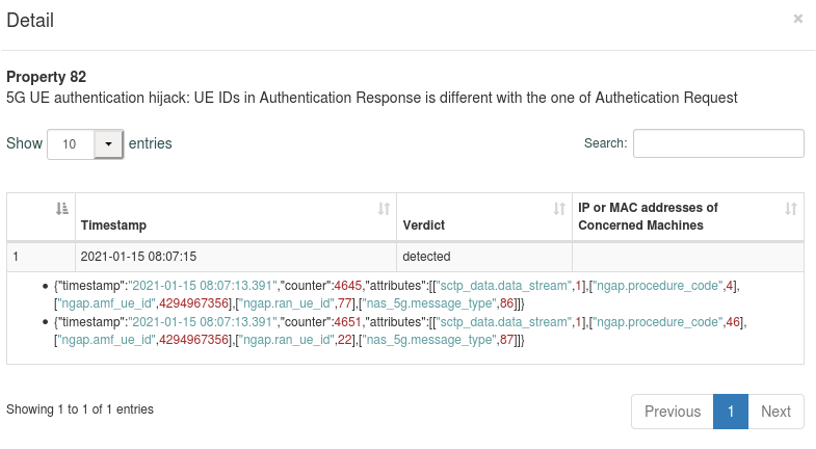}
    \caption{MMT alert when detecting the modified traffic generated by 5Greplay}
    \label{fig:mmt_scenario1}
\end{figure}

The traffic was successfully modified and replayed, and MMT triggered an alert to indicate the suspicious activity in the network. Figure \ref{fig:mmt_scenario1} shows the notified alert, with the modification in the \texttt{ngap.ran\_ue\_id}  attribute of the packet that should be 77 but is 22. From this scenario we can conclude that 5Greplay can modify PCAP files and replay them to a NIC, enabling the evaluation of rule-based IDSs using the fuzz operator \textit{CH\_ATTR(P)}. 

\subsection{Scenario 2: Sending malformed packets against open-source 5G cores in real-time}
\subsubsection{Description}
This use case scenario evaluates the capability of 5Greplay to systematically create and send malformed packets to a 5G core network, in order to evaluate its robustness against unexpected entries at run-time.

\subsubsection{Testbed Setup}
We performed this scenario against the two 5G core open-source solutions, free5GC and open5GS. In both cases we used the RAN simulator UERANSIM. Free5GC and UERANSIM are installed on two different virtual machines, connected via a private network, while open5GS and UERANSIM are installed on the same machine.\par

\begin{figure}
    \centering
    \includegraphics[width=8cm]{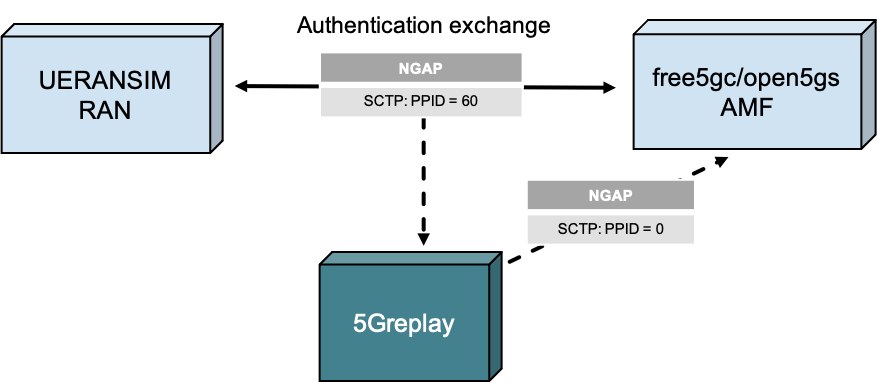}
    \caption{Sending malformed NGAP packets against free5GC}
    \label{fig:scenario2}
\end{figure}

To test the online forwarding functionality of 5Greplay, we configured 5Greplay to detect NGAP protocol messages sent by the UE during the authentication exchange, and replayed them to the AMFs, changing the SCTP protocol identifier from 60 to 0, in such a way that we could evaluate the robustness of the core services against malformed NGAP packets. We depict this scenario in the Figure~\ref{fig:scenario2}.  \par

\begin{figure}
    \centering
    \includegraphics[width=12cm]{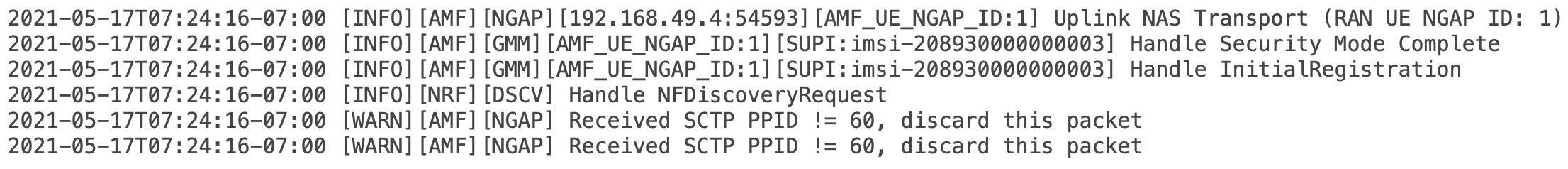}
    \caption{free5GC AMF log when receiving a malformed NGAP packet}
    \label{fig:free5gc_SCTP60}
\end{figure}

\begin{figure}
    \centering
    \includegraphics[width=12cm]{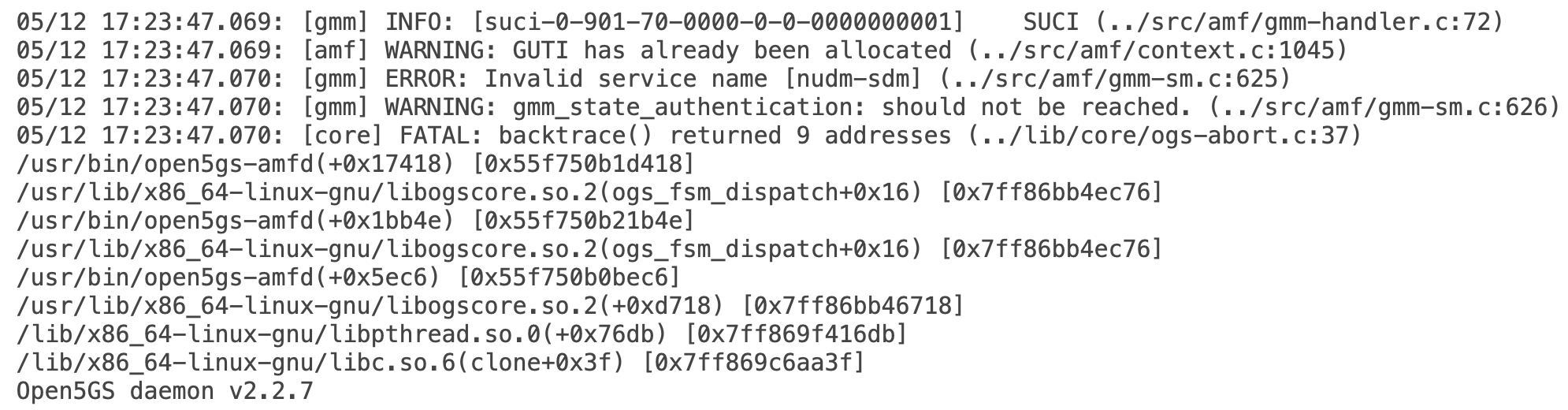}
    \caption{open5GS AMF log when receiving a malformed NGAP packet}
    \label{fig:open5gs_SCTP60}
\end{figure}

\subsubsection{Analysis of the results}

When replaying against free5GC, we got an AMF warning, but the simulator keep running and allowed new UE connections, as showed in the Figure \ref{fig:free5gc_SCTP60}. Therefore, we conclude that it is protected against this type of malformed NGAP packets. On the other hand, open5GS was not able to handle this packet and the simulator crashed, preventing new connections to the AMF, as depicted in Figure \ref{fig:open5gs_SCTP60}.\par

In this case, we evaluated the capability of 5Greplay to replay and modify 5G network packets in a online way, through a private network or in a same machine by using the fuzz operator \textit{CH\_ATTR(P)}. 

\subsection{Scenario 3: NAS-5G SMC Replay attack}
\label{sec:scenario-smc-replay-attack}
\subsubsection{Description} 
This scenario evaluates the use of 5Greplay to perform security tests by modifying and injecting network traffic into a specif target. ENISA, in its threat landscape for 5G networks mentioned before, 
reported that AMFs are vulnerable to replay attacks of the NAS Security mode control (SMC) procedure messages. Moreover, the 3GPP TS33.512 specification for the  AMF Security Assurance\cite{etsi_133512} proposes a test case to verify if an AMF is properly protected against NAS SMC replay attacks. The NAS SMC procedure is initiated by the AMF after a successful authentication of an UE in order to establish a security context that enables encrypted communication between the UE and the AMF. In addition, the AMF could send a NAS SMC message in an already existing security context to change the security algorithm in use.\par

Before the establishment of the security context, the NAS messages are not encrypted. A malicious actor, with access to the NAS traffic in the interface N1, could intercept a NAS SMC \textit{Security Mode command} clear message sent from the AMF to the UE, copy its NAS sequence number (NAS SQN), and use it to build a NAS SMC \textit{Security Mode complete} message that is replayed to the AMF, or directly intercept a NAS SMC \textit{Security Mode complete} message and replay it to the AMF.\par

If the AMF does not implement a proper protection against this type of attack, the network will not drop the replayed packet. Furthermore, a malicious actor could use this technique to impersonate an UE and force the system to reduce the security level by using a weaker security algorithm or turning the security off and, thus, compromise the system. We depict this scenario in the Figure~\ref{fig:scenario1}.  \par

\begin{figure}
    \centering
    \includegraphics[width=8cm]{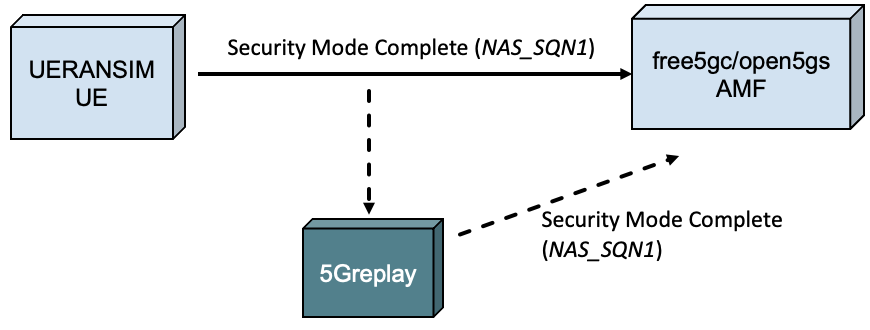}
    \caption{NAS-5G SMC Replay Attack}
    \label{fig:scenario3}
\end{figure}

\subsubsection{Testbed Setup}

We performed this security test scenario in a 5G open-source simulation platform based on free5Gc and UERANSIM. To test the online forwarding functionality of 5Greplay and the protection of the open5GS AMF against replay attacks, we configured the 5Greplay in order to detect the NAS SMC \textit{Security Mode complete} messages sent by the UE after its authentication, and replay it twice to the AMF. Then, we checked the AMF logs, and we monitored the network to verify that the AMF actually received the same packet twice.

\subsubsection{Analysis of the results}

\begin{figure}
    \centering
    \includegraphics[width=12cm]{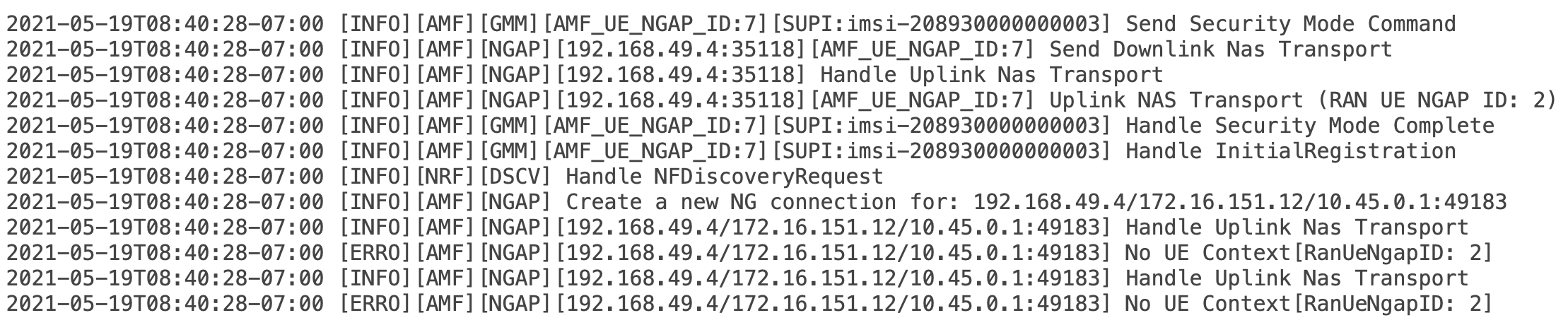}
    \caption{Free5Gc AMF log when replaying Security Mode Complete messages (SMC)}
    \label{fig:SMC_free5gc}
\end{figure}

As depicted in the free5Gc AMF log and shown in Figure \ref{fig:SMC_free5gc}, after the NAS SMC \textit{Security Mode Command} message, the AMF received a legitimate  NAS SMC \textit{Security Mode complete message}, and two NGAP packets with the same UE NGAP ID as the legitimate user. The AMF identified this as not belonging to the same NGAP security context. These two packets corresponded to the replayed packets by 5Greplay and allow us to conclude that the free5Gc AMF is protected again this type of replay attack. Moreover, we proved the utility of 5Greplay to perform standardized security tests by using the fuzz operator \textit{CH\_ATTR(P)}.

\subsection{Scenario 4: High-bandwidth traffic generation}

\subsubsection{Description}
The main objective of this scenarios is to validate the scalability of 5Greplay.For this, we configured the tool to replay the traffic as much as possible. The tool supports both libpcap and Data Plane Development Kit (DPDK). By using only one thread, the tool easily reaches the full-rate of 9.56 Gbps and 1.28 Mpps when testing on a Intel Ethernet Network Adapter X710.
We profile the high performance of packet generation by performing what can be considered DoS attacks or stress tests on the two 5G cores, open5GS and free5GC, that have been setup for the previous scenarios.

\subsubsection{Testbed Setup}

We replay a pre-captured PCAP file against a 5G core service. The PCAP file contains traffic of a complete UE session. To generate such traffic, we firstly start \texttt{nr-gnb} in UERANSIM to activate the gNodeB. We then start \texttt{nr-ue} in UERANSIM to connect to the open5GS. After that, we stop \texttt{nr-ue} and then \texttt{nr-gnb}.
The PCAP file is then used by 5Greplay to inject its packets against the targeted core service. We repeat the test many times. 5Greplay has a parameter \texttt{nb-copies} that indicates the number of copies of a packet that must be created and injected into the network. We increase \texttt{nb-copies} by 10 for each successive replay. 
The test is stopped when we see an error in the execution log of the service.

\begin{lstlisting}[caption={5Greplay rule to forward packets to AMF services}, style=xmlStyle, label=lst:rule-forward-packet]
<property property_id="103" type_property="FORWARD" description="Inject packets from UE to Core">
    <event description="From UE and NAS-5G packets"
           boolean_expression="( ( #is_same_ipv4(ip.src, '127.0.0.1') ) &amp;&amp; ((sctp.dest_port == 38412 ) &amp;&amp;(sctp.ch_type == 0)) )"/>
</property>
\end{lstlisting}

The packets being forwarded are selected by the rule in Listing~\ref{lst:rule-forward-packet}. This is a simple rule having only one event. It selects any packets issued from \texttt{127.0.0.1} that is the IP address of the UERANSIM UE, and the SCTP destination port 38412 that is assigned by IANA to be used by the NGAP protocol\cite{etsi_138412}. The last condition, \texttt{sctp.ch\_type == 0}, allows selecting only the SCTP data chunk, thus avoiding replaying SCTP HEARTBEAT packets that are used to maintain the SCTP connections. 

\subsubsection{Analysis of the results}



\begin{figure}
    \centering
    \includegraphics[width=12cm]{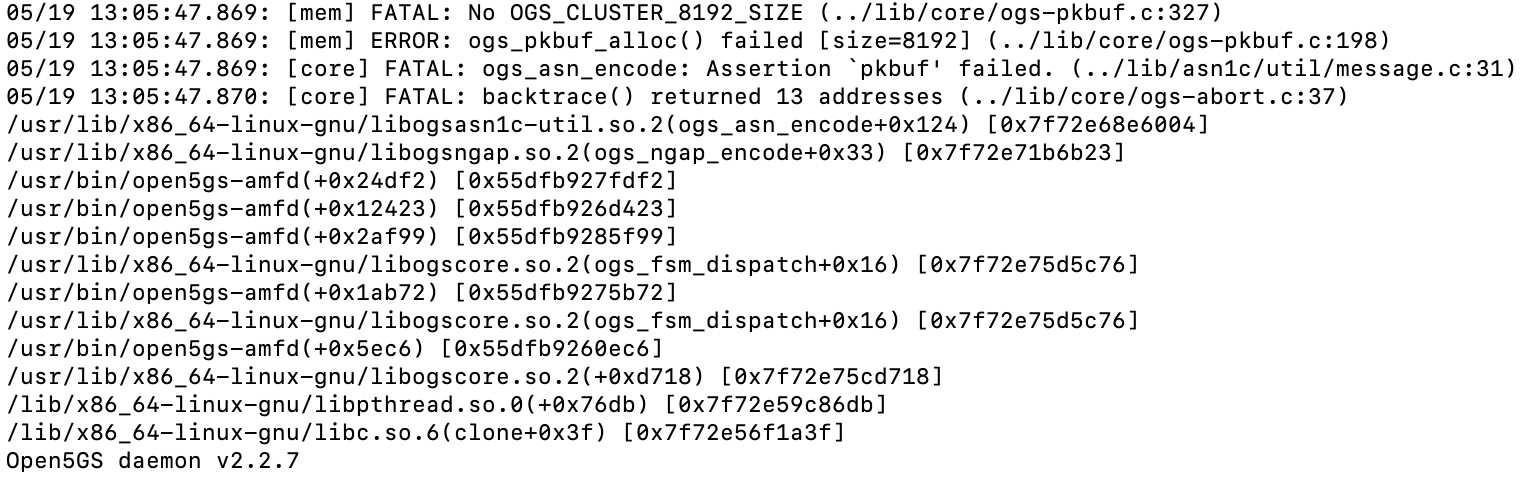}
    \caption{Error in open5GS AMF when replaying 1780 packet copies}
    \label{fig:dos-open5gs}
\end{figure}

Table~\ref{tab:cores-endurance} resumes the tests on AMF services of open5GS and free5GC.
When replaying network traffic against open5GS, we got an \texttt{OGS\_CLUSTER\_8192\_SIZE} error when the number of copies reaches 1780, as  depicted in Figure~\ref{fig:dos-open5gs}. The average replaying rate is about 509.5 pps and 834 kbps.

After looking into the issues tracker of open5GS, we found that open5GS AMF allocates its memory pool to be able to handle by default a maximum of 1024 UEs. We then changed the \texttt{pool} parameter of the AMF to increase the memory pool, so that the AMF can handle more. Nevertheless, AMF crashed again when we increased the \texttt{nb-copies} parameter of 5Greplay.

\begin{table}[h!]
    \centering
    \begin{tabular}{|r|r|r|r|r|}
    \hline
        &  \#packet copies & Avg. packets/s  & Avg. kbit/s    \\ \hline  \hline
open5Gs & 1780      &      509.5   &  834    \\ \hline
free5GC &  3000     &      594.9   &  974  \\ \hline
    \end{tabular}
    \caption{Endurance of 5G AMF services against traffic replaying}
    \label{tab:cores-endurance}
\end{table}

\begin{figure}
    \centering
    \includegraphics[width=12cm]{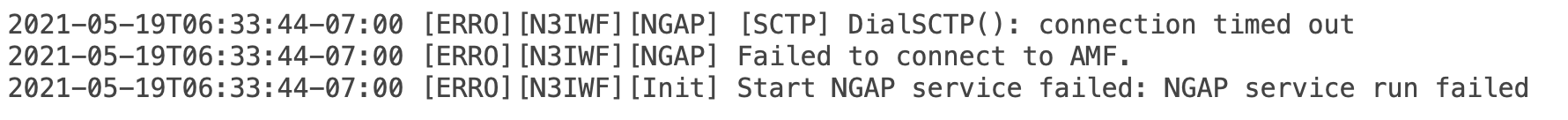}
    \caption{Error in free5GC AMF when replaying 3000 packet copies}
    \label{fig:dos-free5gc}
\end{figure}

When evaluating free5GC we got the errors showed in Figure~\ref{fig:dos-free5gc} when the number of copies is more than 3000. When the number of copies is exactly 3000, 5Greplay generated an average traffic rate of 594.9 pps and 974 kbps. Since these rates are very low with respect to the ability of 5Greplay when using DPDK, we can conclude that 5Greplay can be used to test the robustness of 5G core services when supporting a huge number of UEs by using in the fuzz operator \textit{DUP\_PKT(P)}.

\section{Conclusion}
\label{sec:conclusion}

In this paper, we have introduced our 5G replay tool to address the lack of an open-source tool to perform security testing in 5G networks. To achieve such purpose, the tool provides a flexible way to modify 5G protocol packets before injecting them to the target services to be tested. We presented in the paper four scenarios to evaluate the tool, including for testing the scalability of the tool. The experimentation results show that the alternated packets are correctly formatted with respect to the 5G standard protocols, such as, NAS-5G, NGAP. These packets are accepted by the 5G core services. The tool even provoked crashes in the targeted services when performing DoS attacks.

For future works, we will extend the tool to take into account the rules that allow users to generate packets from scratch. We are also defining new 5G attacks that can be replayed by the tool, and investigating techniques to manage encrypted traffic. We plan to provide an easy way to replay a complete session, and intend to implement and test new ways to alter packets, such as changing the order of two packets. Furthermore, the experimental evaluation will  be performed on other 5G interfaces.

\begin{acks}
This research is supported by the H2020 projects SANCUS N° 952672, INSPIRE-5Gplus N° 871808, and the French ANR project MOSAICO N° ANR-19-CE25-0012.

\begin{figure}[ht]
    \includegraphics[width=2cm]{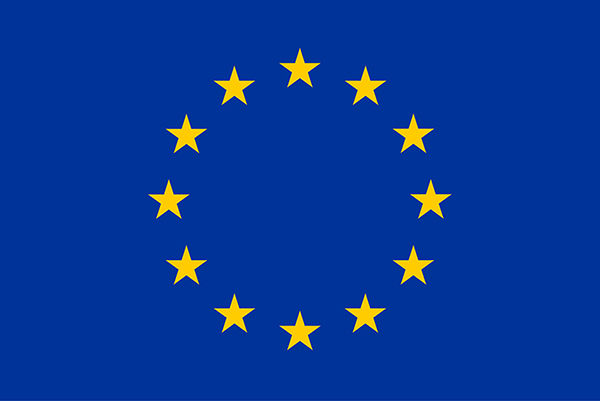}
\end{figure}
\end{acks}

\bibliographystyle{ACM-Reference-Format}
\bibliography{biblio}

\end{document}